\documentclass[11pt,a4paper,notitlepage]{article}
\usepackage{amssymb}
\usepackage{color,graphicx}
\usepackage{amsmath}
\usepackage{bm}
\usepackage{float}
\usepackage{geometry}
\usepackage{amssymb}
\usepackage{float}
\usepackage{amssymb}
\usepackage{color,graphicx}
\usepackage{amsmath}
\usepackage{amsbsy}
\usepackage{amsthm}
\usepackage{bm}
\usepackage{epsfig}
\usepackage{lscape}
\usepackage{float}
\usepackage{authblk}

\begin{document}

\title{Collapse Models: a theoretical, experimental and philosophical review}
\author[1]{Angelo Bassi}
\affil[1]{
Department of Physics, University of Trieste, Strada Costiera 11, 34151 Trieste, Italy}
\affil[1]{Istituto Nazionale di Fisica Nucleare, Trieste Section, Via Valerio 2, 34127 Trieste, Italy}
\author[2]{Mauro Dorato}
\affil[2]{Department of Philosophy, Communication and Media Studies, Universit\`a degli Studi Roma Tre,
Via Ostiense 234, 00146, Rome, Italy}
\author[3]{Hendrik Ulbricht}
\affil[3]{School of Physics and Astronomy, University of Southampton,Southampton SO17 1BJ, United Kingdom}
\date{}
\maketitle

\begin{abstract}
In this paper, we review and connect the three essential conditions needed by the collapse model to achieve a complete and exact formulation, namely the theoretical, the experimental, and the ontological ones. These features correspond to the three parts of the paper. In any empirical science, the first two features are obviously connected but, as is well known, among the different formulations and interpretations of non-relativistic quantum mechanics, only collapse models, as the paper well illustrates with a richness of details, have experimental consequences. Finally, we show that a clarification of the ontological intimations of collapse models is needed for at least three reasons: (1) to respond to the indispensable task of answering the question `what are collapse models (and in general any physical theory) about?'; (2) to achieve a deeper understanding of their different formulations; (3) to enlarge the panorama of possible readings of a theory, which historically has often played a fundamental heuristic role.
\end{abstract}

\section{Introduction}

One of the main tasks of any physical theory is to try to answer the question: \textit{what is the theory about?}" It is usually said that this interpretative task is rather easy except for quantum mechanics and that the conceptual problems of the latter theory, are not eased by considering its relativistic generalization. In fact, this judgment needs qualification.  Two examples will suffice.  In the case of Newtonian mechanics, the lack of direct observability of the gravitation \textit{force} has always caused some conceptual perplexities. Not by chance, until the end of the 19th century the great physicist Heinrich Hertz ~\cite{Her89} tried to formulate classical mechanics by disposing of the concept of force: all we can observe directly are accelerations. The same question can be raised about Classical Electromagnetism, our second example: do lines of force exist? In his valuable ~\cite{Lan} Marc Lange has answered this question in the negative.
The need for an interpretation of the mathematical formalism in which a physical theory is formulated is important not only from a philosophical viewpoint. An attempt to clarify the ontology of a physical theory has often had heuristic value. At the end of the XIX century, there was a widespread disagreement among famous physicists about the ontology of our best physical theories: some considered the atomistic hypothesis to be only a useful fiction (Ostwald, Mach and Poincar\`{e} among others), and others (like Boltzmann) firmly believed in the mind-independence existence of atoms. Spurred by this controversy, physicists attempted to solve the dispute \emph{experimentally}: the convergence of 13 experimentally different ways to calculate Avogadro's number became a clear piece of evidence in favor of the existence of atoms (~\cite{sal84}). The whole community of physicists quickly converted to the hypothesis that atoms were \emph{real or mind-independent}. Consequently, the energetist approach, according to which the primary stuff in nature is energy, was progressively abandoned. In our case, if we don't try to construe mathematically precise models providing a satisfactory, exact answer to the crucial question: \emph{"why and how does a measure of the position of an electron that, when going through two slits, was in a state of superposition of positions turns non-locally into a well-localized point on a fluorescent screen?} we will probably miss important future developments of theoretical physics obtainable by new experiments. 
Furthermore, there seems to be a sort of sociological change in the physics community. Few authoritative figures of contemporary physics (among which Gell-Mann and Weinberg, both winners of the Nobel prize for physics) have changed their minds about the importance of a deeper study of the conceptual foundations of contemporary physics. This also implies going back to the founding fathers' philosophical discussions ~\cite{BV}. By following the seminal work by John Bell, Roger Penrose, Hugh Everett and GianCarlo Ghirardi among others, contemporary physicists are beginning to realize that the measurement problem, besides its intrinsic interest, may even be a stumbling block toward the construction of a quantum theory of gravity. An attentive philosophical analysis is therefore called for.

First of all, as Maudlin has remarked ~\cite{MAU}, a theory is not realist or instrumentalist \textit{per se}. To be a realist or instrumentalist is to have an attitude toward a theory. Instrumentalists in general argue that the aim of science is to construe empirically adequate theories, so that the function of physics is to predict and control the physical world. Realists, on the contrary, claim that the aim of science is to provide a consistent description of a mind-independent physical world and of our place in it. They often regard the instrumentalist attitude toward quantum theory -  often unawarely absorbed by students in the physics departments - as unreasonable: quantum theory as standardly taught, they claim, is "not even a theory", precisely because of its lack of ontological clarity, or \textit{exactness} as Bell put it.  Even the staunchest defender of a realistic stance toward the theory,  however,  must accept the claim that merely instrumentalist attitudes (of the type shut up and calculate) toward the quantum state cannot be viewed as inconsistent, since, as suggested above, they call into play the overarching \textit{aim} of the scientific enterprise.  
 
In this paper, we will take a realistic attitude without further justification. As is well known, the main candidates for a realistic approach toward quantum mechanics are (1) Everett's/many worlds', (2) Bohmian mechanics' and (3) collapse models'. As mentioned above, in our review we will focus on (3) by offering a novel, multiperspectival approach, bringing together not only the theoretical and experimental aspects but also a philosophical discussion of the main conceptual problems presented by the first two aspects. In this sense, we claim that a combination of these three aspects can offer a more complete and therefore a deeper understanding of the current, relevant literature.
Before beginning, however, an extremely brief comparison with the other two realistic approaches is appropriate
(i)
The ontology of Everettian quantum mechanics is about the \textit{wave function}. By postulating the splitting of the physical world in any interaction, this view changes the metaphysics without changing the physical theory. There is no dualism of evolution described by Schr\"{o}dinger equation on the one hand,and by a non-linear irreversible dynamics described by the Born rule. The measurement problem is thereby solved.
On the contrary Bohmian mechanics and dynamical collapse models are \textit{different} theories and not, strictly speaking, an \textit{interpretation} of quantum mechanics. 
(ii)
Bohmian mechanics \textit{adds} to Schr\"{o}dinger equation a so-called 'guiding equation', which specifies the velocities of the particles in terms of the wave function. In particular, the velocities of each particle depend non-locally on the positions of all the others. In the Bohmian case, the ontology of the theory is essentially one of \textit{particles}, while the status of the wave function, allegedly evolving in configuration space, is more controversial. In any case, Bohmian mechanics is presented as a \textit{completion} of the standard theory, which instead presupposes the two evolutions but is regarded as complete.
(iii)
In the case of dynamical collapse models  Schr\"{o}dinger equation is \textit{modified or better generalized} via the addition of a non-linear term. In some sense, Schr\"{o}dinger equation is "wrong" and, as we will see, needs to be supplemented in an appropriate way. The ontology of this theory is more pluralistic than the other two, consisting in the hypothesis that the wave function denotes either (i) in a galaxies of events as Bell put it (known as \textit{flashes}) being the result of localizations of the wave function or (ii) in a matter field propagating in configuration space\ (iii) in the 3N-dimensional configuration space itself describing the N particles composing the physical system or (iv) a  dispositional property of an ensemble of particles, which is the hypothesis that we will defend.

Within the collapse theories, the first two ontologies are based on spatiotemporally extended "be\textit{ables}", a term introduced by Bell in opposition to observa\textit{ble}). The third ontology is about the 3N configuration space, which describes the configuration of the particles of the system. The fourth involves primitive dispositions \textit{quantitative propensities} possessed by the particles to localize  In a word, as Bell put it, if we want to formulate quantum mechanics \textit{exactly}, the wave function must either be incomplete (Bohm) or not always right (GRW). 
The following review of collapse models consists in a \textit{synthesis} of three different aspects, namely a \textit{theoretical}, an \textit{experimental} \textit{and} a \textit{philosophical} one, at a level that is technically more advanced than, say, some among the many books available in the literature (\cite{MAU}, ~\cite {GHI}, ~\cite{ALB}).  In this sense, we claim that its value consists in synthetising three different but inseparable dimensions of the collapse models that should have always been discussed together. We believe that such a synthesis may provide a deeper understanding of one of the main research programs in the foundations of physics.

In the first part of the paper, we briefly summarize the main theoretical features of the collapse models. In the second part, we present possible experimental tests of the theory. In the last part , we evaluate the three above-mentioned ontological assumptions (flashes, matter density, configuration space) by evaluating them in view of the first two parts of the paper

\section{The GRW model} \label{sec:two}
As is well-known, the key problem of quantum theory is how to reconcile the quantum nature of the microscopic constituents of matter with the classical properties of composite systems such as macroscopic objects. The textbook formulation of the theory ultimately assumes a mysterious division between the microscopic quantum world and the macroscopic classical one, but why there is such a division, and where it precisely lies, is not explained. The theory only says that when performing a measurement which connects the micro and the macro world, the quantum wave function collapses to a definite state. But again, why it collapses and when it does so, is not spelled out in clear terms.   

Collapse models~\cite{Grw,Csl,Pr,rep2} aim at solving this problem by combining the linear and deterministic quantum dynamics and the collapse of the wave function, which is nonlinear and stochastic, into a single dynamical equation, capable of accounting for the quantum properties of protons, neutrons and electrons, the classical properties of macroscopic object, as well as the (smooth) transition from one domain to the other.  As a matter of fact, the title of the original GRW paper~\cite{Grw} is {\it  Unified dynamics for microscopic and macroscopic systems.}

Collapse models assume that any physical system, be it large or small, is ultimately quantum mechanical and as such it is described by a wave function. The time evolution of the wave function is guided by a dynamical equation, which departs from Schr\"odinger dynamics. Precisely, it is assumed that every constituent of the physical system is subject to {\it spontaneous collapses} in space. They occur at {\it random times} and are governed by a given probability law, characterized by the {\it collapse rate} $\lambda$, i.e. the number of collapses per unit time. In mathematical terms, what happens during a collapse is that the wave function $|\psi_t \rangle$ of the whole system changes instantaneously to 
\begin{equation} 
|\psi_t  \rangle  \rightarrow  \frac{L_i(\bold{x}) |\psi_t \rangle}{ \|L_i(\bold{x}) |\psi_t \rangle \|},
\end{equation}
 $L_i$ the localization operator defined as
\begin{equation} \label{eq:co}
L_i(\bold{x})= \left(\frac{\alpha}{\pi}\right)^{\frac{4}{3}}\exp \left(- \frac{\alpha}{2}(\hat{\bold{q}_i}-\bold{x})^2 \right),
\end{equation}, $\hat{\bold{q}_i}$ is the position operator of the $i$-th constituent of the system suffering the collapse, and $\bold{x}$ the center of the collapse. We see that a collapse corresponds to multiplying the global wave function by a Gaussian function (and normalizing again the state), which suppresses those parts of the wave function that are far away from the center ${\bf x}$ of the collapse, and keeping only those that are close to the center: the wave function is localized in space, with a precision controlled by the length $r_c=1/\sqrt{\alpha}$. 

For example, if the wave function {\it before} the collapse is in a superposition of states which are distant more that $r_c$, a collapse suppresses one of the two terms and amplifies  the other, so that {\it after} the collapse the wave function is localized in space (again, with respect to the resolution set by $r_c$). On the other hand, if the wave function is in a superposition of states that are closer than $r_c$, the collapse will be ineffective as none of the terms will be suppressed. This is an important feature, because a collapse which is too sharp in space will jeopardize the internal structure of matter, according to which the wave function of electrons can be delocalized over several atoms. For this reason, the suggested~\cite{Grw} numerical value of the precision of the collapse is $r_c \sim 10^{-7}$m, which is a typical mesoscopic distance, meaning with this that macroscopic superpositions are suppressed, while microscopic ones are not. 

The probability for particles $i$ to experience  a localization around a point $\bold{x}$ of space is given by:
\begin{equation}\label{bornrule}
p_i(\bold{x})=\|L_i(\bold{x}) |\psi_t \rangle\|^2,
\end{equation}
which implies that the collapses are more likely to occur around those points in space where the wave function is appreciably different from zero. This is another way of saying that the collapse occurs following (almost) the Born rule. 

In between collapses the wave function evolves following the Schr\"odinger equation. As such, the collapse evolution is piece-wise: the wave function spreads out in space as dictated by the usual quantum dynamics, and enjoys being in a superposition, until a collapse occurs, which localizes it in space; then it can spread out again till the next collapse, and so on. 

The spontaneous collapses must be rare for microscopic systems, otherwise they would have already been spotted. For this reason GRW~\cite{Grw} suggested that $\lambda \sim 10^{-16}$s$^{-1}$ meaning that for a single electron or proton a collapse occurs on average once every $\sim 10^8$ years, which is more or less never. Then for microscopic systems the new dynamics is almost identical to the usual quantum dynamics, modulo tiny deviations, which in light of potential tests of these models are the most interesting ones. 
What makes collapse models interesting and viable as a consistent quantum theory is the inbuilt amplification mechanism: when one particle of a composite system suffers a localization, the wave function of the {\it entire} system is localized. Therefore, if we take for example a typical macroscopic object with $10^{24}$ constituents, each of which is subject to a collapse once every $\sim 10^{16}$s, the wave function of the system suffers a localization once every $\sim 10^{-8}$s: every second there are about $10^{8}$ collapse occurring somewhere in the system, which keep its wave function well localized in space. 

Then the picture which emerges is the following: the wave function of microscopic systems is spread out in space:  Schr\"odinger equation, which makes it diffuse via the kinetic term, or creates superpositions through the interaction term. Collapses are so rare, that they can be safely neglected. When particles interact to form more complex systems, their wave functions become entangled in a unique global wave function, which is subject to the collapses associated to its constituents. In this situation, the amplification mechanism enters into play: the collapse rate associated to the system's wave function is proportional to the number of its constituents, and if this number is large enough, as for macroscopic objects, the collapse rate becomes high enough to guarantee that the system's wave function has no time to spread out in space. As a consequence, the wave function of a macroscopic object is always well-localized in space, so well-localized that it behaves, for all practical purposes, like a point, moving subject to external forces, as for Newtonian mechanics.

According to the previous analysis, the solution to the quantum measurement problem offered by collapse models is the following: a microscopic system enters the measurement process in a quantum state that can be in any superposition allowed by the experimental situation; if no external influence disturbs it, this superposition is stable,  During the measurement process (here we follow the ideal scheme proposed by von Neumann), the microscopic system becomes entangled with the macroscopic apparatus, and the superposition starts propagating from the first to the second. However, before this happens, the many collapses occurring in the macroscopic instrument destroy the superposition state: the instrument will always be in a localized state, corresponding to one of the possible outcomes, and also the state of the microscopic system will be one of those that previously were in a superposition, in particular dthe one correlated to the outcome of the measurement. In additions, it can be proven that the states to which the superposition can collapse are distributed according to the Born rule.  

This is how the GRW model, and collapse models in general, account for measurement situations in quantum theory and, equivalently, for Schr\"odinger's cat paradox:  the  classical world of tables and chairs (and cats) emerges from a wavy quantum world via a nonlinear dynamics, according to which wave functions become the better localized in space, the more constituents are glued to each other.

As a note, it should be clear that, within the context of collapse models, it is inappropriate to speak of microscopic `particles', having in mind localized objects. We did, and will use this term, as it is customary among physicists, but here it is misleading. At the microscopic level, the wave function of a `particle' is spread out over space, and it would be odd to associate it to a point-like object. Only at the macroscopic level wave functions are well-localized and can be associated to localized objects, although in an approximative sense. 

Two comments are in order. The first one is that the collapses as described before do not preserve the symmetry (or anti-symmetry) of the wave function representing identical particles. For example, this implies that electrons in an atom would slowly all decay to the ground state because of the collapses. The stability of matter is thus jeopardized. This issues can be resolved by replacing the collapse operator defined in Eq.~\eqref{eq:co} with a suitable operator preserving the identity of particles. 

The second comment is that the piece-wise dynamics previously outlined, although consistent, might look somehow artificial. This issue can also be resolved by introducing a continuous version of the collapse, which acts on the wave function alongside with the Schr\"odinger's. The resulting dynamics is a diffusion process for the wave function in the Hilbert space, with the unitary part or the collapse part dominating, depending on the size of the system, i.e. on the number of its constituents. 

The model thus outlined, where the collapse is continuous and preserves the identity of particles, is called Continuous Spontaneous Localization (CSL) model~\cite{Csl}.

\section{Experimental tests of collapse models}

It is clear from the previous discussion that compared to quantum mechanics, collapse models make different
predictions. 
In this section, we discuss about possible experiments proposed
to test collapse models against standard quantum theory. For a more detailed analysis of this topic, we refer to~\cite{ad1,sci,rep2,tumulka}. 

The most intuitive test of collapse models is realized by testing spatial quantum superpositions of massive objects that are much heavier than, say, a Cesium atom. For given parameters, collapse models forbid superpositions that are instead perfectly allowed by standard quantum theory. The simple idea behind the experimental test is to generate such superpositions state and tune the parameters so that the collapse effect would become apparent - if existing. Qualitatively, the parameters that one needs to control are the mass $m$ of the particle, the spatial size of the superposition state $l$ and the time for the superposition to exist $t$. The direct link of $[m, l, t]$ with collapse model's parameters has to be defined for each individual experimental configuration and model. In CSL[Adler], such a comparison has been done in~\cite{rep2} at the level of Lindbald's master equation. This helps to identify the appropriate parameters for the test. 

We note that this implies that experiments only test collapse models at the level of the density matrix, which is the \textit{same} for decoherence effects. However, the decoherence approach, in contrast to collapse models, does not discuss the reason for wavefunction collapse as an intrinsic property of the dynamics of the quantum system~\cite{Zur, Joo}. Decoherence is triggered by interactions of the quantum system with its environment. Experimental studies - at the density matrix level - confirm indeed the existence of decoherence~\cite{HaroNob, WinNob, Ar2, Horn1}. This means that both collapse models and decoherence predict very similar effects to observe in the density matrix dynamics and great care has to be taken to distinguish the two effects.

Before we discuss in a bit more detail the experimental tests of collapse models, we want to mention the so-called {\it macroscopicity} measures. The quest to test collapse models naturally aims to test quantum effects in the macroscopic domain.  The definition of 'marcoscopic' must be handled with case, as the most prominent collapse models scale with the mass of the quantum object. This means that a useful marcoscopicity measure must include \textit{mass} as a parameter. Other quantum systems, such as entangled photon pairs, are known to exist over many kilometers [Zeilinger 144km], but would not represent a good system to test collapse models as the rest-mass of the photons is notoriously zero. Even if the objective choice for a measure of macroscpicity is still lively debated, we think it appropriate to mention the measure $\mu$ by Nimmrichter and Hornberger~\cite{Nim}, which fits in well with the test of collapse models by matter-wave-interferometry-like experiments - as it combines the set of parameters [$m, l, t$]. With $\mu$ at hand, it is possible to objectively compare very different experiments ranging from optomechanics to superconducting flux devices and how effective they are to test collapse models. 

As mentioned above, the kinds of experiments invoked to directly test spatial superposition of massive particles is based on matter-wave interferometry and the largest particles that are supposed to show interference are \textit{organic molecules}. Experiments are done in the Talbot-Lau interferometer (TLI) configuration, which acts in favour of the low coherence in molecular and nanoparticle beams~\cite{Juf}. The largest particles interfered so far are of mass 10$^4$~amu in the Vienna interferometers~\cite{Bra, Fei}. This experiment is still about two orders of magnitude too small to test the CSL model with the strongest bound according to Adler, given the mass of the record molecule, the separation of slits on the gratings of the TLIs, which is about 100~nm and the time for particles to travel through the TLIs, which is on the order of 1~ms. 

Interestingly, by assuming the sense of macroscopicity given by $\mu$, the generation of a superposition between the ground and the excited state of a very massive mechanical harmonic oscillator, such as that prepared in quantum optomechanics~\cite{Asp} cannot easily outperform the lighter molecules in the Talbot-Lau interferometers. The reason is that all three parameters [$l, m, t$] have to be sufficiently large. In the case of optomechanical systems $m$=10$^{14}$~amu and is therefore much larger than for molecules, but the spatial separation between the states in a superposition is very small; more precisely, if estimated with the spatial size of the zero-point motion of the mechanical harmonic oscillator, it is of the order of $l$=10$^{-15}$~m or less, i

Also atoms are not likely to produce much stronger tests. Atom interferometry of the Mach-Zehnder type can be performed with very large beam separations~\cite{Kov}, which gives makes parameter $l$ of the order of $cm$ and therefore very competitive for a test, or can be hold for as long as 20 seconds in a dipole trap~\cite{Xu}, which makes the parameter $t$ quite large. However, the mass of a single atom is so much lighter compared to molecules in TLIs, that the macroscopicity is not larger. If those atoms would undergo interference together, a collection of atoms would of course improve the mass factor,  However, all interferometry experiments of Bose-Einstein condensates of atoms have revealed the individual superposition of atoms and not of the condensate ensemble, the superposition state being a N-particle product state of single-atom superpositions instead of a NOON state, which is the case of a molecule. The mass of the object in superposition can easily be worked out from interference fringe pattern and this would be the individual atom. If (i) ultra-cold atoms could be bound together or (ii) many of them could be made interacting with each other, say in a dipolar fashion, (iii) and we could realize a matter-wave interferometer for atoms, we would make a very good test of collapse models. Developments are underway and cold atomic interferometers hold promise for a test as this quantum technology is rich and powerful in preparation of massive quantum states. 

Future plans for matter-wave interferometers which achieve even larger macroscopicity involve the OTIMA and LUMI interferometers at Vienna~\cite{Has, Fei}. OTIMA and LUMI are Talbot-Lau Interferometer and should be able to interfere particles of masses up to 10$^7$~amu and beyond. The machine is exiting, and now intense beams of slow nanoparticles have to be generated. 

Experiments with single-levitated nano- and micro-particles have been suggested in 2010~\cite{Bar, Ori, Cha}. More details for such an experiment have been worked out for a double slit type of experiment in free fall~\cite{Rom}. A related scheme for a nanoparticle based on a Talbot interferometer has been proposed as well [our paper], while interferometry of levitated nanoparticles has yet to be achieved. Experiments with a particle of 10~nm diameter [l=100nm, t=2ms] would directly test the CSL model with Adler's parameter choice~\cite{Bel}. Investigations of intrinsic noise in optically levitated systems naturally lead to the suggestion of ion Paul trapping of charged particles or even better the magnetic levitation of superconductors and a rapidly growing experimental community is taking care of levitated nano- and microparticles in vacuum~\cite{Gon}. Ground-state cooling as a first step toward macroscopic quantum systems based on levitation has recently been achieved ~\cite{Del, Mag, Teb}. Matterwave interferometry with levitated mechanical systems comes into experimental reach and many alternatives for realisation do exist. for a recent comprehensive review, see ~\cite{Mil}.

All such matter-wave test have common limitations. Known decoherence effects due to collisions with other particles such as rest gas in vacuum chambers or interaction [emission, scattering, absorption] of thermal radiation make these experiments a massive technical challenge. Furthermore, all particles have to propagate freely, which means that they are affected by Earth's gravity. Particles of different mass fall by the same distance if in free motion for the same time. The interferometry of more massive particles takes more time compared to interferometry of lighter particles. This means that more massive particles have to be in free fall for longer and fall for more time. To keep the matter-wave experiment in reasonable dimensions [say 10~m in the vertical direction], the mass of the particle cannot be much larger than 10$^7$~amu, c.f.~\cite{Bel}. Experiments in micro-gravity in space would allow overcoming this limit. The collaboration MAQRO to work towards a space mission has been formed a decade ago and is preparing large-mass matterwave interferometry in space~\cite{Gas, Kal}. Alternatively, a test beyond 10$^7$~amu could be done on Earth, if the particle could be prepared in a sufficient macroscopic superposition while trapped at low noise and possibly the evolution time of the wavefunction can be accelerated or boosted by an inflation operation~\cite{Ori2, Mar}.

However, better upper bounds on $\lambda$ can be obtained by studying
the indirect effects of the collapse. For example, collapse models predict an increase of the kinetic energy of any system, leading to consequences at cosmological
scale. More precisely, this increase of energy implies the heating of the intergalactic
medium (IGM) that amounts to a change on its spectrum. This effect
was studied in~\cite{ad1} setting an upper bound $\lambda\leq10^{-9}\:\textrm{s}^{-1}$. 

Nowadays, the best upper bound on the $\lambda$ parameter comes from
the study of the process of spontaneous radiation emission. In fact, as shown
for the first time in~\cite{fu} and later discussed more in detail
in~\cite{ar,bd,abd,dirk,basdon,basdon2}, the interaction with the collapse noise induces, for charged
systems, an emission of radiation not predicted by standard quantum
mechanics. The radiation emitted from a single particle is very small,
but when a macroscopic object containing a number of atoms of the
order $10^{23}$ is considered, the effect becomes much more relevant.
A comparison between the radiation emission predicted by the CSL model
with experimental data coming from the emission from Ge
has been done in~\cite{fu}, leading to an upper bound $\lambda\leq10^{-11}\:\textrm{s}^{-1}$. A recent dedicated experiment at Gran Sasso has lead to stronger bounds~\cite{Don}.

For completeness, in table~\ref{exptest} we report all the bounds
on $\lambda$ coming from different experiments and cosmological data~\cite{rep2}:

\begin{table}[h]
\centering
  \begin{tabular}{| l| c| }
    \hline
Experiments and cosmological data & Upper bound on $\lambda$ (s$^{-1}$) \\ \hline\hline
Matter-wave interferometry& $10^{-5}$\\ \hline
Decay of supercurrents (SQUIDS) & $10^{-3}$\\ \hline 
Spontaneous X-ray emission from Ge & $10^{-11}$\\ \hline 
Proton decay & $10$\\ \hline 
Dissociation of cosmic hydrogen & $1$\\ \hline
Heating of intergalactic medium (IGM) & $10^{-9}$\\ \hline
Heating of interstellar dust grains & $10^{-2}$\\ \hline

     \end{tabular}
\caption{Upper bounds on the CSL parameter $\lambda$ coming from laboratory experiments (first four lines) as well as from the analysis of cosmological data (last three lines).}\label{exptest}
 \end{table}

The technical complications for the implementation of marcoscopic laboratory table-top matter-wave experiments has lead to a number of proposals for alternative, among which non-interferometric tests of quantum superposition. Such ideas began with the treatment of the collapse field as a noise field that, if the given experiment would be sufficiently sensitive, could in some way be observed as a distortion. The first of such ideas consisted in the proposal to discuss CSL effects for the light-matter interaction in a generic two-level system, as widely discussed in quantum optics. Then the CSL noise is just another noise which effects the emission properties of an excited two-level system. The CSL noise triggered the relaxation of the exited electron and therefore the emission of radiation sooner than expected from the natural lifetime at the exited state. The result is a shift and broadening of the related spectral emission line as worked out in detail in Bahrami et al.~\cite{Bah1}. Unfortunately both effects from CSL and other collapse models are very small and ultrahigh precision spectroscopy experiments are still at least two orders of magnitude away to resolve the effect on real spectra. The two-level effect does not include the N particle amplification mechanism as the spontaneous photo-emission process explained above. A further advantage of the spontaneous emission effect is that it predicts a new spectral feature (to trigger a forbidden transition and to stimulate its emission) instead of a modification (shift, broadening) of an existing spectral line. To prove or disprove the existence of a new feature is the preferred situation for an experimental test. 

A related non-interferometric test - and one which has been already done with existing experimental quantum technology~\cite{Bar} - is a test with optomechanical and magnetomechanical systems. In optomechanics, a mechanical harmonic oscillator is coupled to light. A typical setting is cavity optomechanics, where the mechanical oscillator is a nanofabricated cantilever, which forms one of the mirrors of an optical resonator, setting the scene for an optical light mode. The optical cavity mode now depends on the motion of the mechanical oscillator. Light of a given wavelength will be enhanced in the cavity not depending on the position of the cantilever. The properties of the mechanical oscillator are mapped on the spectral response of the light field, which makes an ultra-sensitive mechanical sensor. Such optomechanical systems have been pioneered in the past decade or so, and many experiments already exist~\cite{Asp, Bar}.  CSL noise now affects the motion of the cantilever, which consists of many atoms. The light reads this effect of the CSL noise, which can be described as a heating effect on an initially cooled centre of mass motion of the cantilever. This results in an increase in the area of the spectral line which corresponds to this motion. The heating effect depends on shape and size of the mechanical object. This very same effects can also be predicted for the aforementioned levitated mechanical systems and it is interesting to note that the very same CSL's heating effect is also predicted for the direct observation of the motion/rotation/vibration of suspended or levitated objects. In sum, mechanical systems have a high potentiality for ultimate testing of CSL, not only by interferometric, but also by non-interferometric means and are covering a large parameter space for testing CSL models. It should be noted that those mentioned above are just a small selection of possible non-interferometric experiments for testing collapse models. For a comprehensive review we refer the reader to the recent work by Carlesso et al.~\cite{Cal}.

This variety of ideas lay ought to be considered as future, very promising  tests of collapse models while giving, at the same time the opportunity to work with fascinating experiments towards the understanding of nature and especially of quantum systems in the macroscopic regime.

\section{The physical origin of the collapse }

The random localization processes in the GRW model and the noise field in the CSL model are postulated, but an explanation of their origin in physical terms is still lacking The collapse is introduced so to speak 'by hand' in order to arrive at a phenomenological description of the wave function's collapse, with the correct quantum probabilities given by the Born rule. The origin of the collapse, however, is still an open question. In this section we discuss  two possible answers about the origin of the collapse that  have been proposed in the literature.

One option, championed by  Di{\'o}si and Penrose, is that the spontaneous collapses  have a gravitational origin~\cite{Pen, qmupl, diosi1}. 
Penrose~\cite{Pen} has studied the effects of gravity on a superposed state $\left|\psi\right\rangle =a\left|\psi_{1}\right\rangle +b\left|\psi_{2}\right\rangle $
where $\left|\psi_{1}\right\rangle $ and $\left|\psi_{2}\right\rangle $
are two identical stationary states corresponding to the same energy eigenvalue, one located around a given region ``1" and the other one located around another region ``2" of spacetime. According to standard quantum mechanics, this superposition state is also stationary and therefore should last forever. On the other hand, the two states are located in different positions and therefore correspond to two different matter densities. According to General
Relativity, this means that they curve spacetime in different ways. Then a superposition of different spacetime geometries appears, which amounts to an ill-defined time translation operator. This, according to Penrose, leads to an uncertainty in energy of the superposed state $\left|\psi\right\rangle$, which, in the Newtonian limit, should
be proportional to the gravitational self-attraction of the two superposed
states, i.e.:
\[
\Delta E=-4\pi G\int\int d\mathbf{x}\, d\mathbf{y}\,\frac{\left(M_{1}\left(\mathbf{x}\right)-M_{2}\left(\mathbf{x}\right)\right)\left(M_{1}\left(\mathbf{y}\right)-M_{2}\left(\mathbf{y}\right)\right)}{|\mathbf{x}-\mathbf{y}|},
\]
where $M_{1(2)}$ corresponds to the mass density distribution of the state $\left|\psi_{1(2)}\right\rangle $. 
This uncertainty in the energy might suggests that the superposed state
$\left|\psi\right\rangle $ collapses in a time of the
order $\tau\sim\hbar/\Delta E$. 

Not only does Penrose's argument imply that that wave function should collapse, and that the larger the object in the superposition state, the greater the gravitational self-attraction $\Delta E$ and the faster the collapse; but it also provides a quantitative estimation of the reduction time. Clearly, this argument is heuristic and does not provide a dynamical equation
for the evolution of the wave function. 

The gravity-induced collapse model proposed by Di{\'o}si~\cite{qmupl, diosi1} provides instead such a dynamics by postulating that the state vector evolves as the CSL model mentioned in Section~\ref{sec:two}, with a different choice for the collapse operator, in such a way that the collapse time coincides with that indicated by Penrose. Hence the name Di{\'o}si-Penrose (DP) model. The CLS model and DP model thus are equivalent at the formal level, including the amplification mechanism; by using a different choice for the collapse operator, they differ on the quantitative level. As discussed before, the virtue of the DP model is to suggest a physical origin for the collapse, although, as also suggested by Penrose, the ultimate answer should come from a proper quantum theory if gravity.

Another possibility to understand the origin of the noise is to consider
collapse models not as fundamental theories, but as phenomenological description of a deeper underlying theory. ``Trace dynamics'', proposed by Adler~\cite{trace}, gives such a possibility. In trace dynamics, the dynamical variables are non-commutative matrices (operators) whose dynamics is described in terms of a trace Lagrangian and a trace Hamiltonian, which are generalization of the standard Lagrangian and Hamiltonian formalism. It is assumed that this highly non trivial dynamics is very fast, and rapidly reaches statistical equilibrium; the resulting dynamics for the canonical ensemble takes the form of quantum field theory. In other words, quantum theory emerges as a thermodynamic limit of the underlying trace dynamics. 

The fluctuations from this thermodynamics limit takes the form of a Brownian motion's corrections to the dynamics, which eventually produces stochastic modifications to standard quantum theory. The hypothesis, yet to be proven, is that the Brownian motion' corrections are exactly the same as those described by collapse models. Therefore, in trace dynamics, the collapse of the wave function should arise from the underlying dynamics. 
The research for an underlying theory is not yet concluded. In fact, as pointed out by Adler itself in the introduction of his book, ``[...] while we have given a general framework in which an emergent quantum theory may appear, we have not identified a specific theory in which all our requirements are realized''~\cite{trace}.

\section{Non locality and the problem with relativistic generalizations}

We conclude with a discussion about non locality and relativistic collapse models. Non locality, in standard quantum theory, appears because the collapse of the wave function is nonlocal; this is required by the fact otherwise it would not be possible to explain quantum correlations for entangled systems. Similarly, in collapse models, the collapse induces non local effects: given an entangled state of two systems, the collapse of one of the two systems instantaneously affects the state of the other . Moreover,
as in standard quantum theory, the randomness of the collapse makes it impossible to use these instantaneous effects to send faster than light signals. Bell immediately recognized this fact and some other features of the GRW model: ``I am particularly struck by the fact that the model is as Lorentz invariant as it could be in the non relativistic version''~\cite{bell}. 

On the one hand, insofar as collapse models are considered as merely phenomenological, there
is no need to require them to be Lorentz invariant. 

On the other hand, if collapse equations are taken as fundamental
, then a Lorentz invariant description should be preferred. Different attempts were carried out in this direction. The most direct
approach is to consider a Lorentz invariant version of the CSL model~\cite{rel1}.
In order to guarantee Lorentz invariance, the noise correlation is postulated to be $\mathbb{E}\left[w(x)w(y)\right]=\delta^{(4)}(x-y)$
with $x, y$ denoting two different spacetime points. 
However, this implies a noise which is delta correlated in space, leading to 
an infinite increase of energy, thus making the model physically inconsistent. The natural way out is to replace the white noise with some colored Lorentz
invariant noise. However, this leads to serious nonlocal features in the
dynamical evolution of the states, which makes difficult any further progress with this approach . Other attempts to arrive at a relativistic collapse model were carried out by Pearle~\cite{rel2}, Dowker and Henson~\cite{rel3}, Tumulka~\cite{rel4} and Bedingham~\cite{rel5}. 

\section{Which ontology for collapse models?}

The main problem that collapse models try to solve is to clarify the unclear notion of measurement, on which the standard view is based what are the physical conditions necessary and sufficient for a measurement interaction to occur? Using an argument \textit{ex auctoritate} we can motivate this question by this quotation by John Bell, stressing once more the role of "aims" in the realist/instrumentalist divide \begin{quote}
"experiment is a tool. The \textit{aim} remains: to understand the world. To restrict quantum mechanics to be exclusively about piddling laboratory operations is to betray the great enterprise. A serious formulation will not exclude the big world outside the laboratory." \end{quote}(my emphasis)
By recalling the three main ontological hypotheses posed by collapse models, we will begin by discussing the "wave function" interpretation, First, we may want to remark that the wave function, qua a (complex) \textit{function} on the configuration space, is an abstract entity. For the sake of precision, one could follow Maudlin's suggestion and replace the ambiguous term 'wave function' with \textit{quantum state}, where the latter is purposedly to be interpreted as a physical entity \textit{denoted} by the wave function regarded as an abstract object. However,  once we are clear about the need to distinguish the wave function regarded as an abstract entity (a function) from its denotatum, our main problem of course still remains, which is to figure out what Maudlin's quantum state \textit{is}. 
In what follows, and keeping in mind Maudlin's caveat, in accord with the literature we will keep using the word 'wave function' given that it appropriately suggests that it can be identified as an entity evolving in configuration space: No one can understand this theory until he is willing to think of the wave function as a real objective field rather than just a probability amplitude
even though it propagates not in 3-space but in 3N-space (\cite{Bell}, p. 128).
To the extent that the wave function becomes the fundamental \textit{physical entity} presupposed by quantum mechanics, it seems natural to assume that the 3N configuration space on which it propagates is promoted to the role of being the fundamental physical space for all non-standard formulations of the theory, collapse models included, given that they describe the world in terms of the wave function too~\cite{ALBI,Ney}. Since the wave function is defined in configuration space, realism about the wave function may be misleadingly suggested, among other things, by the fact that in GRW models, the collapse is referred to by \textit{mathematical} \textit{multiplication} of the global wave function by a Gaussian. Here we will argue that if we want to make sense of the ontology of collapse models, this is the wrong way to go. In such models, wave function realism must entail that the multiplication in question \textit{represents} a \textit{field} taking values at points in configuration space, not in physical space
The main problem with this position is to explain the \textit{emergence} of the familiar, macroscopic world of tables, chairs and people from a 3N dimensional space, which is one of the main tasks which must be accomplished by a theory that aims at "mending" in part the standard formulation. Arguments in favor and against the viability of this explanatory research program have been thoroughly discussed in~\cite{Ney}) to which we refer the reader. Here we want to make \textit{two} critical points.
The \textit{first} is that, in our opinion, no convincing explanation has been provided so far, the main difficulty depending on the vague notion of \textit{emergence}. In philosophy, this concept is used extensively, and not only in the philosophy of physics, to the detriment of clarity. As far as we can tell, the only clear meaning to be associated to 'emergence' is \textit{reduction}, which, in turn, by following Nagel and Hartmann, ought to be regarded as a \textit{deduction} of the emergent properties from those characterizing the relevant fundamental level. The paradigmatic example is here the logical relation between thermodynamic and statistical mechanics. The property of the former 'level' (i.e. hot and cold for example referred to a gas) emerge from the statistico-mechanical description in \textit{this} sense. 
Of course, one cannot deny the possibility of a reduction meant in this sense, as this denial might certainly be due to our lack of imagination.  However,  it must be admitted that, so far, no such a deduction is available or has been proposed and talking about the emergence of the familiar macroscopic world from a physically fundamental configuration space is a mere (though worth pursuing) research program. The \textit{second} critical point is a consequence of the first, since lacking this reduction or at least an exact account of 'emergence', it is not clear how wave function realism can really solve the measurement problem, since it would always be not be clear what it means to claim that a three-dimensional pointer in a physical space has moved to the right.\footnote{For this point, see \cite{ALLDORZANLAU}, p.375.}  

We can then go on discussing the so-called primitive ontology hypotheses calling into play local beables (entities) living in Newtonian \textit{spacetime}. The explanatory task referred to above becomes easier, since macroscopic objects live in 3D space. In the literature one typically finds two interpretations of the `hits' of the wave function, that is two localizations hypothesis, namely in terms of a galaxy of flashes GRW$_{f}$ in Bell's metaphor, and of matter density GRW$_{m}$. 
In the former hypothesis, a table or a chair is constituted by 
pointlike 'flashes', representing the center of the peak of the Gaussian, each of them evolving in time in a discrete, non-continuous way. Since these localization processes happen in three-dimensional space at a certain moment of time, and objects are constituted by microscopic components, we must regard the former as literally constructed out of these flashes. It must be admitted that this hypothesis looks at odds, or maybe just ad hoc, with what we know about the microstructure of these objects. By looking 'ad odds' we don't mean that it contradicts well-known, solid facts about chemistry like 'water is made of $H_2 O$ but that it is still not \textit{conceptually} connected with them. It just looks as an addition to such facts that it is left somewhat unexplained. As Maudlin puts it
``there is literally nothing at all material that is localized in spacetime" ~\cite{MAU} (p. 115). To this remark, we add that the main lesson of quantum field theory is that wave and particles are oscillations of quantum fields, so in our most fundamental physical theory, fields must be regarded as more fundamental than parts-less particles like electrons.

A more reasonable hypothesis is that the 'hits' of the wave function refer to the ``density-of-stuff" ontology proposed by Ghirardi et al. ~\cite{Grw}.  
In this version of the collapse models, the mass density of the fundamental field constituting physical reality suffers contractions in correspondence to the multiplication of wave function in configuration space, which, as in the flash ontology, must be are regarded as mathematical tools to describe the concrete evolution of the matter field in ordinary space and time. Unlike flashes, which are discrete, pointlike happenings in spacetime, a field is obviously both continuous and smeared in space like a field must be, so that it occupies a non-infinitesimally extended region of space. A 'hit' in the abstract space of the wave function describes a nonlocal change in how much of the physical field is (its density) in a certain region. 
More precisely, the ontology in this version of the collapse model is given by a scalar field $\rho$ in three-dimensional space, that is, an assignment of values to each of its points \textbf{r} at a certain time t. This assignment clearly depends on the values of the wave function in configuration space. 
Following the notation by (\cite{ALLDORZANLAU}, p.375) we can write the field $\rho$ as
 $\rho(\textbf{r},t)$ =  $\sum_{i} m_{i} \rho_{i}(r,t)$ where  $\rho_{i}(r,t)$ stands for the density of each particle$i$, and $m_{i}$ stand for its mass.

For instance, a matter-field crossing the two slits will localize at a point in the second screen by increasing \textit{almost all} of its density there, the qualification `almost all' being necessary for the so-called `tail problem' ~\cite{ALLO}. The problem in question refers to the tails of the Gaussian multiplying the wave function, which contain some very small amount of matter, which does not make any difference at the empirical level. As Maudlin points out, another advantage of the matter density ontology over the discrete flash ontology, is its coherence with the CSL models described above (\cite{MAU}, p.121)
There is an important aspect of the collapse model that so far we have neglected: the localizations (flashes or density of matter) are irreversible, stochastic processes happening \textit{in} spacetime. As is well-known, all the fundamental physical laws are time-symmetric, that is, they describe the processes unfolding in what we conventionally call past to the future direction as well the time-reversed process going in the opposite direction (future to past). Albert ~\cite{AL} and other eminent scholars have proposed to regard the collapse of the wave function (in those views in which it is postulated) as an irreversible process justifying our belief in a time asymmetry physical world. The localizations featuring essentially in the collapse models could then be regarded as a more precise, lawlike explanation of macroscopic \textit{de facto} irreversible phenomena like the entropic growth in closed systems or the prevalence of retarded vs advanced radiation. 
By endowing the matter field in three-dimensional space with irreducibly stochastic, spontaneous and irreversible \textit{dispositions to localize} ~\cite{DOES}, the probabilities of these localizations could be assigned numbers by introducing a propensity-based interpretation of probability. In this alternative account, the 'hits' of the wave function would denote a set of irreducible dispositions of the matter field to manifest in changes in its density. It should be remarked that, in all collapse views of quantum mechanics, a dispositionalist ontology must presuppose that there is nothing that 'triggers' the disposition; that is, there is no \textit{stimulus} that, like a match or a stone, that causes the match to burn or the glass to break. Whether the presence of stimuli is essential for a disposition to manifest in an event is a problem that here cannot be discussed.

\section{Conclusions}

In this paper, we have tried to connect the three essential conditions for a complete and overall formulation of a physical theory, namely the \emph{theoretical}, the \emph{experimental} and the \emph{philosophical/ontological} ones. As required by any formulation of a physical theory, the first two features are connected since, as is well known, among the different formulations and interpretations of quantum mechanics, GRW is the only approach that has an experimental dimension because, as illustrated above, is falsifiable. The ontological, philosophical component is also needed for two reasons: 1) it responds to the indispensable \textit{scientific} task of answering the question 'what is a physical theory about? Precisely for these reasons, 2) it helps to better understand the different theoretical formulations of a physical theory. In our context, for instance, the two theories GRW$_{f}$  GRW$_{f}$ have different conceptual links with the CSL research program. As such, not only is an exact ontology an indispensable requirement of any physical theory, but enlarging the panorama of possible reading of a theory may play a fundamental heuristic value.

\section{Acknowledgements}
A.B. acknowledges financial support from the EIC Pathfinder project QuCoM (GA no. 101046973), the PNRR PE National Quantum Science and Technology Institute (PE0000023), the University of Trieste and INFN


\begin{thebibliography}{86}

\bibitem {Her89} Herzt, H. Die Prinzipien der Mechanik in neuem Zusammenhange dargestellt Leipzig: Johann Ambrosius Barth 1894

\bibitem {Lan} M. Lange. The Philosophy of Physics, Blackwell, 2002.

\bibitem {sal84} W. Salmon \textit{Scientific Explanation and the Causal Structure of the World}, Princeton University Press (1984)

\bibitem{BV} Bacciagaluppi G. and Valentini A. (2009) (eds) Quantum Theory at the Crossroads: Reconsidering the 1927 Solvay Conference, Cambridge University Press.

\bibitem {MAU} T. Maudlin,  Philosophy of Physics.  Quantum theory. Princeton University Press (2019)

\bibitem{GHI} G.C. Ghirardi, Sneaking a Look at God's Cards: Unraveling the Mysteries of Quantum Mechanics - Revised Edition. Princeton University Press (2007)

\bibitem{ALB} D. Albert, Quantum Mechanics,  and Experience.  Harvard University Press (1992)

\bibitem{AL} D.Albert, Time and Chance, Harvard University Press, (2003)

\bibitem{Grw}
G. C. Ghirardi, A. Rimini, and T. Weber, {Phys. Rev. D} {\bf 34}, 470-491 (1986).


\bibitem {ALBI} D. Albert, After Physics. Cambridge MA, Harvard University Press (2016)

\bibitem {Ney} The wave function (eds A. Ney and D. Albert). Oxford
University Press, 2013.

\bibitem {ALLDORZANLAU} V.Allori, M. Dorato, F. Laudisa, N. Zangh\`{i}, La natura delle cose, Carocci, Roma, (2006)

\bibitem {ALLO} A. David, B. Lower, The tails of Schr\"{o}dinger cat in Perspectives on Quantum Reality, edited by R. Clifton, Kluwer, p.81-92 (1996).

\bibitem{Bell1987}
J. S. Bell, {Schr\"{o}dinger:Centenary Celebration of a Polymath.}, Cambridge University Press (1987).

\bibitem{Breuer}
H.-P. Breuer, F. Petruccione, {The Theory of Open Quantum Systems}, Oxford University Press (2007).

\bibitem{Csl}
G. C. Ghirardi, P. Pearle, and A. Rimini, {Phys. Rev. A} {\bf 42}, 78 (1990).


\bibitem{Cslmass}
G. C. Ghirardi, R. Grassi, and F. Benatti, {Found. Phys.} {\bf 25}, 5-38 (1995).

\bibitem{ad1} S.L. Adler,  J. Phys. A {\bf 40}, 2935 (2007). Ibid. A {\bf 40}, 13501 (2007).

\bibitem{BBU}
M. Bahrami, A. Bassi, and H. Ulbricht, {Phys. Rev. A} {\bf 89}, 032127 (2014).

\bibitem{BB}
M. Bahrami, M. Paternostro, A. Bassi, and H. Ulbricht, {Phys. Rev. Lett.} {\bf 112}, 210404 (2014)

\bibitem{Bell}
J. S. Bell, {\it Speakable and Unspeakable in Quantum Mechanics}, Cambridge University Press (1986).(SPECIFY CHAPTER)

\bibitem{Pe1}
P. Pearle, {Phys. Rev. D} {\bf 13}, 857-868 (1976). 

\bibitem{Pe2}
P. Pearle, {Phys. Rev. A} {\bf 39}, 2277-89 (1989).

\bibitem{Di1}
L. Di\'osi,{J. Phys. A: Math. Gen.} {\bf 21}, 2885-2898 (1988).

\bibitem{Di2}
L. Di\'osi, {Phys. Lett. A} {\bf 129}, 419-423 (1988).

\bibitem{qmupl}
L. Di\'osi, {Phys. Rev. A} \textbf{40}, 1165-1174 (1989).

\bibitem{Pr}
A. Bassi, and G. C. Ghirardi, {Phys. Rep.} {\bf 379}, 257-426 (2003).


\bibitem{sci} S.L. Adler and A. Bassi, Science {\bf 325}, 275 (2009).

\bibitem{rep2} A. Bassi, K. Lochan, S. Satin, T. P. Singh and H. Ulbricht, Rev. Mod. Phys. \textbf{85}, 471-527 (2013).

\bibitem{tumulka} W. Feldmann, R. Tumulka, J. Phys. A: Math. Theor. {\bf 45}, 065304 (2012) .

\bibitem{Ar1} M. Arndt, O. Nairz, J. Vos-Andreae, C. Keller, G. van der Zouw and A. Zeilinger, Nature {\bf 401}, 680 (1999).

\bibitem{Ar2} L. Hackerm\"uller, K. Hornberger, B. Brezger, A. Zeilinger and M. Arndt, Nature {\bf 427}, 711 (2004).

\bibitem{Has} Haslinger, P., Doerre,  N., Geyer, P., Rodewald, J., Nimmrichter, S. and Arndt, M., Nature Physics, {\bf9}, 144-148 (2013).

\bibitem{Horn1} Hornberger K, Gerlich S, Haslinger P, Nimmrichter S, and Arndt M. Rev. Mod. Phys. {\bf84}, 157 (2012).

\bibitem {Per} Jean Perrin J., Les atomes, Librairie Felix Alcan 1913

\bibitem{HaroNob} Haroche, S., Rev. Mod. Phys. {\bf85}, 1083 (2013).

\bibitem{WinNob} Wineland, D.J., Rev. Mod. Phys. {\bf85}, 1103 (2013).

\bibitem{Zur} Zurek, W.H.,  Rev. Mod. Phys. {\bf75}, 715 (2003).

\bibitem{Joo} Joos, E., Zeh, H.D., Kiefer, C., Giulini, D.J., Kupsch, J. and Stamatescu, I.O., Decoherence and the appearance of a classical world in quantum theory. Springer (2013).

\bibitem{Nim} Nimmrichter, S. and Hornberger, K., Phys. Rev. Lett. {\bf110}, 160403 (2013).

\bibitem{Juf} Juffmann, T., Ulbricht, H. and Arndt, M., Rep. Prog. Phys.{\bf76}, 086402 (2013).

\bibitem{Bra} Brand, C., Eibenberger, S., Sezer, U. and Arndt, M., arXiv:1703.02129. (2017).

\bibitem{Fei}  Fein, Y.Y., Geyer, P., Zwick, P., Kiaka, F., Pedalino, S., Mayor, M., Gerlich, S. and Arndt, M., Nat. Phys. {\bf15}, 1242-1245 (2019.).

\bibitem{Asp} Aspelmeyer, M., Kippenberg, T.J. and Marquardt, F., Rev. Mod. Phys. {\bf86}, 1391 (2014.).

\bibitem{Kov} Kovachy, T., Asenbaum, P., Overstreet, C., Donnelly, C.A., Dickerson, S.M., Sugarbaker, A., Hogan, J.M. and Kasevich, M.A., Nature {\bf528}, 530-533 (2015).

\bibitem{Xu} Xu, V., Jaffe, M., Panda, C.D., Kristensen, S.L., Clark, L.W. and Mueller, H., Science {\bf366}, 745-749 (2019).

\bibitem{Gas} Gasbarri, G., Belenchia, A., Carlesso, M., Donadi, S., Bassi, A., Kaltenbaek, R., Paternostro, M. and Ulbricht, H., Commun. Phys., {\bf4}, 1-13 (2021).

\bibitem{Kal} Kaltenbaek, R., Arndt, M., Aspelmeyer, M., Barker, P.F., Bassi, A., Bateman, J., Belenchia, A., Berge, J., Bose, S., Braxmaier, C. and Christophe, B., et.al., arXiv preprint arXiv:2202.01535 (2022).

\bibitem{Bel} Belenchia, A., Gasbarri, G., Kaltenbaek, R., Ulbricht, H. and Paternostro, M., Phys. Rev. {\bf A 100}, 033813 (2019).

\bibitem{Cal}  Carlesso, M., Donadi, S., Ferialdi, L., Paternostro, M., Ulbricht, H. and Bassi, A., Nature Physics, {\bf18}, 243250  (2022).

\bibitem{Mil} Millen, J. and Stickler, B.A., 2020. Contemporary Physics, {\bf61}, 155-168 (2020).

\bibitem{Gon} Gonzalez-Ballestero, C., Aspelmeyer, M., Novotny, L., Quidant, R. and Romero-Isart, O., Science {\bf374}, 3027 (2021). 

\bibitem{Bar} Barker, P.F. and Shneider, M.N., Phys. Rev. {\bf A 81}, 023826 (2010). 

\bibitem{Ori} Romero-Isart, O., Juan, M.L., Quidant, R. and Cirac, J.I., New Journal of Physics, {\bf12}, 033015 (2010). 

\bibitem{Cha} Chang, D.E., Regal, C.A., Papp, S.B., Wilson, D.J., Ye, J., Painter, O., Kimble, H.J. and Zoller, P., Proceedings of the National Academy of Sciences, {\bf107}, 1005 (2010).

\bibitem{Rom} Romero-Isart, O., Pflanzer, A.C., Blaser, F., Kaltenbaek, R., Kiesel, N., Aspelmeyer, M. and Cirac, J.I., Phys. Rev. Lett., {\bf107}, 020405. (2011).

\bibitem{Del} Deli, U., Reisenbauer, M., Dare, K., Grass, D., Vuleti, V., Kiesel, N. and Aspelmeyer, M.,  Science, {\bf367}, 892-895 (2020).

\bibitem{Mag} Magrini, L., Rosenzweig, P., Bach, C., Deutschmann-Olek, A., Hofer, S.G., Hong, S., Kiesel, N., Kugi, A. and Aspelmeyer, M., Nature {\bf595}, 373-377 (2021).

\bibitem{Teb} Tebbenjohanns, F., Mattana, M.L., Rossi, M., Frimmer, M. and Novotny, L., Nature, {\bf595}, 378-382 (2021).

\bibitem{Ori2} Romero-Isart, O., New Journal of Physics {\bf19}, 123029 (2017). 

\bibitem{Mar} Margalit, Y., Dobkowski, O., Zhou, Z., Amit, O., Japha, Y., Moukouri, S., Rohrlich, D., Mazumdar, A., Bose, S., Henkel, C. and Folman, R., Science Advances {\bf7}, 2879 (2021). 

\bibitem{Don} Donadi, S., Piscicchia, K., Curceanu, C., Diosi, L., Laubenstein, M. and Bassi, A., Nature Physics {\bf17}, 74 (2021).

\bibitem{Bah1} Bahrami, M., Bassi, A. and Ulbricht, H., Phys. Rev. {\bf A 89}, 032127 (2014).

\bibitem{Bar} Barzanjeh, S., Xuereb, A., Gr\"{o}blacher, S., Paternostro, M., Regal, C.A. and Weig, E.M., Nature Physics {\bf18}, 1524 (2022).


\bibitem{Ar3} S. Gerlich, S. Eibenberger, M. Tomandl, S. Nimmrichter, K. Hornberger, P. J. Fagan, J. T\"uxen, M. Mayor and M. Arndt, Nature Comm. {\bf 2}, 263 (2011).

\bibitem{fu} Q. Fu, { Phys. Rev.} A {\bf 56}, 1806 (1997).

\bibitem{ar} S.L. Adler and F.M. Ramazanoglu,  { J. Phys. A} {\bf 40}, 13395 (2007); {\it J. Phys. A} {\bf 42}, 109801 (2009). 

\bibitem{bd} A. Bassi and D. D\"urr, { J. Phys. A} {\bf 42}, 485302 (2009).

\bibitem{abd} S.L. Adler, A. Bassi and S. Donadi,  \textit{J. Phys. A} \textbf{46}, 245304 (2013).

\bibitem{dirk} S. Donadi, A. Bassi, D.-A. Deckert, Annals of Physics {\bf 340}, Issue 1, 70-86 (2014). 

\bibitem{basdon} A. Bassi, S. Donadi, Phys. Lett. A {\bf 378}, 761-765 (2014). 

\bibitem{basdon2} S. Donadi, A. Bassi, Journ. Phys. A {\bf 48}, 035305 (2015). 

\bibitem{Pen} R. Penrose, Gen. Relat. Gravit. {\bf 28}, 581-600 (1996).

\bibitem{diosi1} L. Di{\'o}si, Phys. Lett. A {\bf 120}, 377 (1987).

\bibitem{ghi} G.C. Ghirardi, R. Grassi, and A. Rimini, Phys. Rev. A {\bf 42}, 1057 (1990).

\bibitem{bah} M. Bahrami, A. Smirne, A. Bassi, arXiv preprint arXiv:1408.6460, (2014).

\bibitem{trace} S. L. Adler, {\it Quantum Theory as an Emergent Phenomenon}, Cambridge University Press (2004).

\bibitem{bell}J. S. Bell, ``Speakable and unspeakable in quantum mechanics" Cambridge, UK: Cambridge University Press, (2004), chapter 22: ``Are there quantum jumps?".

\bibitem{rel1} G. C. Ghirardi, R. Grassi, and P. Pearle, Found. Phys. {\bf 20}, 1271 (1990).

\bibitem{rel2} P. Pearle, Phys. Rev. A {\bf 59}, 80 (1999).

\bibitem{rel3} F. Dowker, and J. Henson, Journal of Statistical Physics {\bf 115}, 1327 (2004).

\bibitem{rel4} R. Tumulka, Journal of Statistical Physics {\bf 125}, 821 (2006).

\bibitem{rel5} D. J. Bedingham, Found. Phys. {\bf 41}, 686 (2011).


\bibitem {ghi95} G. Ghirardi, R. Grassi, and F. Benatti, Describing the macroscopic world: Closing the circle within the dynamical reduction program?. \textit{Foundations of Physics} 25 , pp. 5-38 (1995).



\bibitem{Alney} Alissa Ney and David Albert (eds.), The wave function, Oxford University Press, 2008.

\bibitem{DOES} M. Dorato, M. Esfeld. Grw as an ontology of dispositions, \textit{Studies in History and Philosophy of Modern Physics} 41, pp. 41-49 (2010).



\end{thebibliography}
\end{document}